%% file: main.tex
\documentclass[12pt]{article}

\usepackage[letterpaper, margin=1in]{geometry}
\usepackage[utf8]{inputenc}
\usepackage{times}
\usepackage{graphicx}
\usepackage{mathptmx}
\usepackage{amsmath}
\usepackage{dsfont}
\usepackage{amsfonts}
\usepackage[justify]{ragged2e}
\usepackage[labelfont=bf,labelsep=period]{caption}
\usepackage[style=nature,maxbibnames=20,url=false,eprint=false,isbn=false]{
    biblatex
}
\usepackage{hyperref}
\hypersetup{
    colorlinks=true,
    linkcolor=blue,
    filecolor=magenta,      
    urlcolor=cyan,
}
\usepackage{titling}
\title{Visualizing near-coexistence of massless Dirac electrons 
and ultra-massive saddle point electrons}
\author{Abhay Kumar Nayak$^{1\dagger}$, Jonathan Reiner$^{1}$, 
Hengxin Tan$^{1}$, Huixia Fu$^{1}$,\\ Henry Ling$^2$, Chandra Shekhar$^{3}$,
Claudia Felser$^{3}$, Tami Pereg-Barnea$^{4}$,\\ Binghai Yan$^{1}$, 
Haim Beidenkopf$^{1\dagger}$, Nurit Avraham$^{1\dagger}$,}
\date{\small $^{1}$ Department of Condensed Matter Physics, 
Weizmann Institute of Science, Rehovot, Israel.\\
$^2$208-5800 Cooney Road, Richmond, British Columbia V6X3A8, Canada
$^{3}$ Max Planck Institute for Chemical Physics of Solids, 
D-01187 Dresden, Germany.\\
$^{4}$ Department of Physics, McGill University, 
Montréal, Québec H3A 2T8, Canada.\\
$^{\dagger}$ Corresponding authors: abhaykumar.nayak@weizmann.ac.il, 
nurit.avraham@weizmann.ac.il, haim.beidenkopf@weizmann.ac.il}

\begin{document} 
%..

\maketitle

\section*{}
\textbf{Strong singularities in the electronic density of states amplify 
correlation effects and play a key role in determining the ordering 
instabilities in various materials. Recently high order van Hove 
singularities (VHSs) with diverging power-law scaling have been classified in 
single-band electron models. We show that the 110 surface of Bismuth exhibits 
high order VHS with an usually high density of states divergence $\sim (E)^{-0.7}$. Detailed mapping of the surface band structure using scanning tunneling microscopy and spectroscopy combined with first-principles calculations show that this singularity occurs in close proximity to Dirac bands located at the center of the surface Brillouin zone. The enhanced power-law 
divergence is shown to originate from the anisotropic flattening of the Dirac 
band just above the Dirac node. Such near-coexistence of massless Dirac 
electrons and ultra-massive saddle points enables to study the interplay of 
high order VHS and Dirac fermions.}

\section*{introduction}
The energy dispersion of the Bloch states govern key physical properties of a 
given material while also regulating the density of states (DOS) which plays a 
pivotal role in determining the ground state of interacting particles. 
Several exotic quantum phases, such as, superconductivity 
\cite{Kohn1965,McChesney2010}, charge density wave \cite{Rice1975,Makogon2011}, 
and magnetism \cite{Fleck1997,Ziletti2015} may manifest due to an upsurge in the 
DOS. In two-dimensions, the vanishing electron velocity at a saddle point in 
the energy dispersion gives rise to a logarithmic divergence in the 
DOS \cite{VanHove1953}, called van Hove singularity (VHS). These VHSs are
 usually accompanied by a topological transition (Lifshits transitions) in 
 which the Fermi surface transforms between a hole-like and an electron-like 
 form due to the appearance or collapsing of a pocket in the Fermi surface. 
 When the Fermi level lies in the vicinity of such Van Hove points, 
 electron-electron interactions are enhanced, 
 promoting the formation of electronic instabilities and correlated states. 
 In addition to ordinary VHS, two-dimensional systems can also host high-order 
 VHS which is characterized by a power-law divergence of DOS in energy. 
 Such high order singularities display more exotic Fermi surface 
 topological transitions around which the dispersion is flatter than near a 
 conventional van Hove point. These stronger divergences in DOS can further 
 enhance the formation of complex quantum phases via interactions and can play 
 important roles in transport phenomena\cite{Yuan2019a,Classen2020}.

High order VHS have been proposed to exist in several electronic systems 
such as twisted bilayer graphene and trilayer graphene, as well as other 
Van der Waals compounds, where the band structure can be tuned by changing the 
twist angle, pressure, or interlayer bias voltage 
\cite{Shtyk2017,Yuan2019a,Guerci2022,Domenico2019,Ramires2012,Bansil2021}. They have been 
associated with the unusual Landau level structure in biased bilayer 
graphene \cite{Shtyk2017}, the nontrivial thermodynamic and transport 
properties in Sr$_3$Ru$_2$O$_7$ \cite{Efremov2019}, the so called supermetal 
with diverging susceptibilities in the absence of long-range order 
\cite{Isobe2019} and more recently with the electronic symmetry breaking in 
CsV$_3$Sb$_5$ \cite{kang2022}.  The effect of a high order VHS on electron 
interaction has been studied theoretically \cite{Yuan2019a,Isobe2019,Classen2020} in a system with weak electron 
interactions. Further theoretical studies classified high-order critical points 
based on topology, scaling, and symmetry, showing that high order VHS can be 
realized at generic or symmetric momenta by tuning a few parameters such as 
twist angle, strain, pressure, and/or 
external fields\cite{Yuan2020,Chandrasekaran2020}. It has been shown that 
they can be obtained by tuning the parameters in the Hamiltonian, 
particularly at high symmetry points in the  Brillouin zone. Nevertheless, 
the corresponding symmetries also restrict the type of singularities 
that can be achieved.  

Bismuth (Bi) is an ideal material to explore the interplay of topological and
correlated phases owing to its high spin-orbit coupling and long mean free path
\cite{Drozdov2014,Schindler2018a,Hsu2019,Jack2019}. Although Bi has been studied
in various contexts, such as unconventional superconductivity
\cite{Prakash2017}, electron fractionalization \cite{Behnia2007} and quantum
hall ferromagnetism \cite{Feldman2016,Randeria2019}, its exact topological
classification has been resolved only recently, as being on the verge of a phase
transition between a high order topological insulator and a strong topological
insulator \cite{Nayak2019,schindler2018}. Here, we show that the (110) surface
of Bi exhibits a remarkable power-law divergence of the DOS arising from
high-order VHS presented on this surface. Spectroscopic mapping of the surface
band structure at the corresponding energy range, using quasi particle
interference (QPI) measurements, shows that this high-order VHS occurs in
conjunction with a highly anisotropic Dirac band located at the center of the
surface Brillouin zone. By directly visualizing the Dirac bands in QPI we
observe a rapid anisotropic flattening of the upper Dirac band just above the
Dirac node which leads to the sharp increase in the DOS. Calculated QPI using
Green's function approach exhibit excellent agreement with our QPI measurements.
The near-coexistence of massless Dirac fermions and heavy saddle-point fermions
is an interesting property of the (110) surface that enables to study the
interplay and competing orders of high orders VHS and Dirac fermions.

\section*{Results} Fresh surfaces of Bi were exposed by cleaving GdPtBi crystals
grown in Bi flux \cite{Shekhar2018,Nayak2019}. Due to its anisotropic
rhombohedral structure cleaving these crystals under ultrahigh vacuum conditions
exposes surfaces of residual crystalline Bi inclusions in various orientations.
Here, we measured the Bi(110) and Bi(111) surfaces (Fig.\ref{fig:Bi110}a), at
4.2K in a low temperature Unisoku STM. A representative topography of pristine
Bi(110) surface terraces is shown in Fig.\ref{fig:Bi110}b and a height profile
across a mono-atomic step edge in Fig.\ref{fig:Bi110}c. A typical differential
conductance ($dI/dV$) profile measured over the Bi(110) surface is shown in
Fig.\ref{fig:Bi110}d. The spectrum was measured on a clean terrace far from any
impurities. Peaks in the $dI/dV$ spectrum (marked by dashed lines in Fig.\ref{fig:Bi110}d) signify increased local density of states (DOS). Many of
them correspond to band extrema shown in the density functional theory (DFT)
calculation of the Bi(110) surface band structure (Fig.\ref{fig:Bi110}e). We consider a crystal of hexagonal shape which preserves $C_3$ rotational and inversion symmetry. The (110) surface preserves only a mirror symmetry. Owing to the inverted bulk band structure of bismuth \cite{Nayak2019,Schindler2018a}, Dirac bands are realized on its various surfaces. While on the Bi(111) surface
the Dirac bands are directly overlapping with bulk bands
\cite{Nayak2019,Drozdov2014,Schindler2018a}, on the (110) surface the Dirac
bands at $\Gamma$ and $M$ (Fig.\ref{fig:Bi110}e) are very well separated from
the bulk projected bands, making it ideal to directly visualize their exact
dispersion.

\subsection*{Visualizing a topological Dirac cone on Bi(110)}
\begin{figure}[ht] \centering \includegraphics[scale=1]{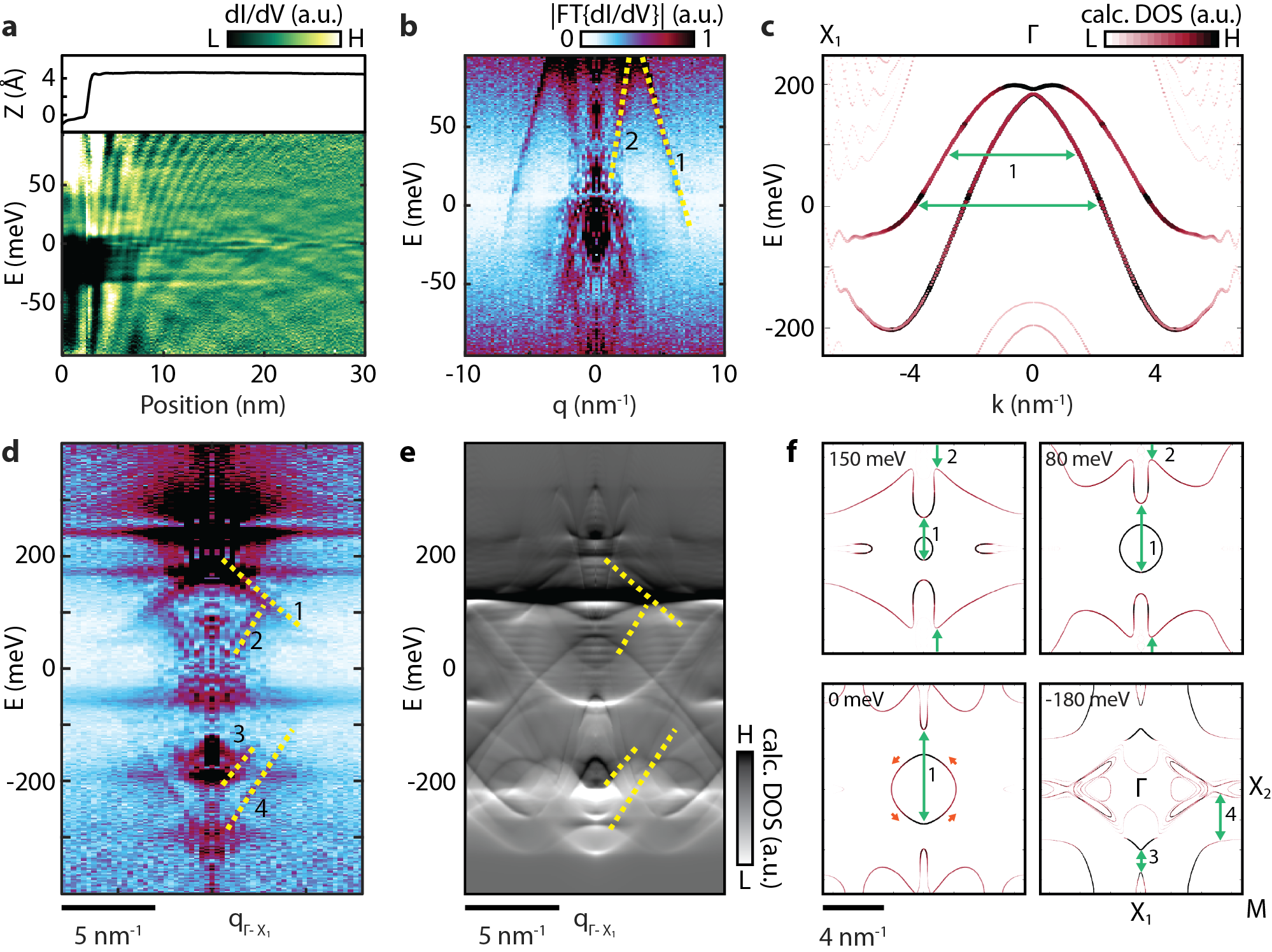}
\caption{\textbf{Quasi-particle interference of a Dirac cone.} \textbf{a,}
Topographic profile along a line perpendicular to the step edge and the
corresponding $dI/dV$ measurement along the line are shown in the upper and the
lower panel, respectively. \textbf{b,} Fourier transform (FT) of \textbf{a}
along the position axis showing the energy dispersion of the scattering wave
vectors (marked by the dashed green lines) along $\Gamma - X_1$. \textbf{c} Ab
initio calculation of the band structure along $\Gamma - X_1$. \textbf{d,}
Fourier transform of a large energy window $dI/dV$ measurement (similar to
\textbf{b}) showing the various dispersing modes. \textbf{e,} Calculated QPI
along $\Gamma-X_1$ using the Green's function approach. The different modes are
marked by dashed green lines in \textbf{c} and \textbf{d}. \textbf{f,} Ab initio
calculation of contours of constant energy (CCE) at different energies for the
Bi(110) surface. The relevant scattering wave vectors are are marked by the
green arrows.} \label{fig:QPI} \end{figure}

To resolve the topological Dirac bands and other surface bands on the Bi(110)
surface we map the QPI patterns imprinted in the local DOS due to scattering off
surface step edges \cite{Avraham2018}. A spatially resolved spectroscopic
$dI/dV$ map measured across an atomic step-edge is shown in Fig.\ref{fig:QPI}a
(see also Fig.\ref{Sfig:QPI_2} and \ref{Sfig:QPI_S2}). The momentum along the
step edge is conserved and we only see scattering processes with momentum
transfer in the direction perpendicular to the step edge. Strong spatial
modulations emanating from the step edge and dispersing in energy are embedded
in the local DOS. Fourier decomposition of these QPI patterns
(Fig.\ref{fig:QPI}b) separates them according to the transferred momentum,
\textbf{q}, between the scattered electronic states \cite{Avraham2018}. We
identify two prominent scattering processes labeled and marked by yellow dashed
lines in Fig.\ref{fig:QPI}b. Particularly interesting is the outer QPI pattern
(process 1) that exhibits clear linear dispersion over more than a hundred meV,
at energies just below the Dirac point at $\Gamma$. This process provides a
direct visualization of the topological Dirac band as it results from scattering
between the lower part of the Dirac cone to its upper part that curves downwards
in a Rashba like manner, as shown in Fig.\ref{fig:QPI}c. A similar larger energy
window measured on different step-edge (Fig.\ref{fig:QPI}d) reproduces the
scattering processes marked in Fig.\ref{fig:QPI}b along with other dispersing
modes at lower energies (see also \ref{Sfig:SP1}, \ref{Sfig:SP2} and
\ref{Sfig:spin}). To fully associate the observed QPI patterns with particular
scattering processes, we compare the data with Green's function calculation of
the QPI (Fig.\ref{fig:QPI}e; see also Fig.\ref{Sfig:GF}), based on \textit{ab
initio} calculation of the Bi(110) surface states. The four identified processes
(yellow dashed lines) exhibit good agreement in \textbf{q} values as a function
of energy with the corresponding calculated QPI patterns.

Further identification of the particular processes is obtained by comparing the
momentum transfer with particular \textbf{q} vectors that connect the
corresponding electron and hole pockets in the calculated surface band
structure. A few representative contours of constant energy (CCE) cuts along
with the identified scattering wave vectors (labeled and marked by green arrows)
is shown in Fig.\ref{fig:QPI}f (see also Fig.\ref{Sfig:SP1} and
Fig.\ref{Sfig:SP2} for details). Sparsity of the surface bands in the relevant
energies allowed us to readily visualize and uniquely identify the various
scattering processes. The scattering process 1 along the $\Gamma - X_1$
direction, indeed corresponds to momentum transfer between the lower part of the
Dirac band, the circular dispersing band around gamma, and its upper part that
curves downwards (Fig.\ref{fig:QPI}c), giving rise to the U-shaped dispersing
pocket. Process 2 results from scattering across the Brillouin zone, between the
high intensity tips of those U-shaped pockets. Finally, processes 3 and 4 result
from scattering between high intensity points along the $\Gamma - X_1$ direction
within the Brillouin zone.

\subsection*{High-order van Hove singularity} 
\begin{figure}[ht] \centering
\includegraphics[width=1\linewidth]{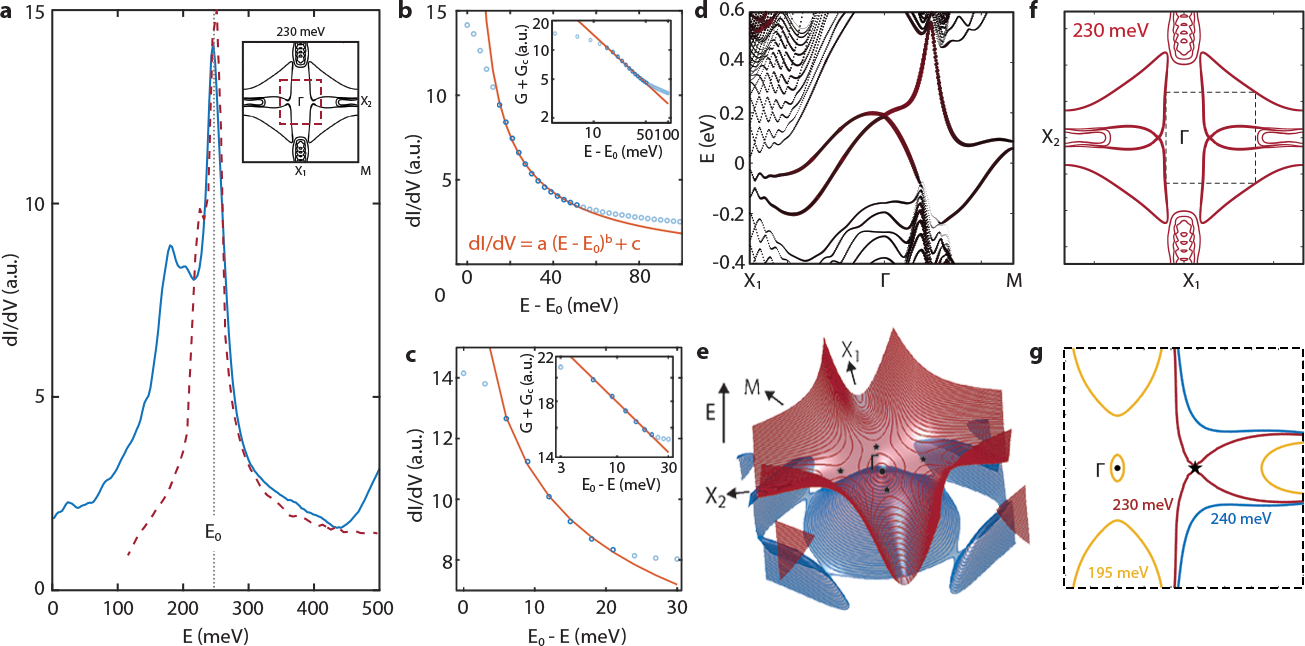} \caption{\textbf{Power-law
divergence in DOS on Bi(110).} \textbf{a,} Zoomed in $dI/dV$ profile around the
$E_0$ = 246 meV peak overlaid with the calculated density of states (DOS)
originating around $\Gamma$ as marked by the dashed box in the inset.
\textbf{b,} $dI/dV$ profile to the right of the peak at $E_0$ extracted from
\textbf{a}. A power-law fit (red) to the raw data (dark blue circle) yields an
exponent of b = -0.7. Inset shows a linear fit (red) to the log-log $dI/dV$
profile, indicating a power-law divergence in DOS due to high-order van Hove
singularity. \textbf{c,} Same as in \textbf{b} for the $dI/dV$ profile to the
left of the peak at $E_0$, yielding a power-law exponent of b=-0.2. \textbf{d,}
Calculated band structure of the Bi(110) surface along a high symmetry line
showing the highly anisotropic Dirac bands. \textbf{e,} Calculated surface band
structure of Bi(110) in the vicinity of the Lifshitz transition. \textbf{f,} The
surface band structure at 230 meV. \textbf{g,} The zoomed in surface band
structure, of the dotted square area in f, in orange, red and black correspond
to 195 meV, 230 meV and 240 meV, respectively.} \label{fig:VHS} \end{figure}

\begin{figure}[ht] \centering
\includegraphics[width=1\linewidth]{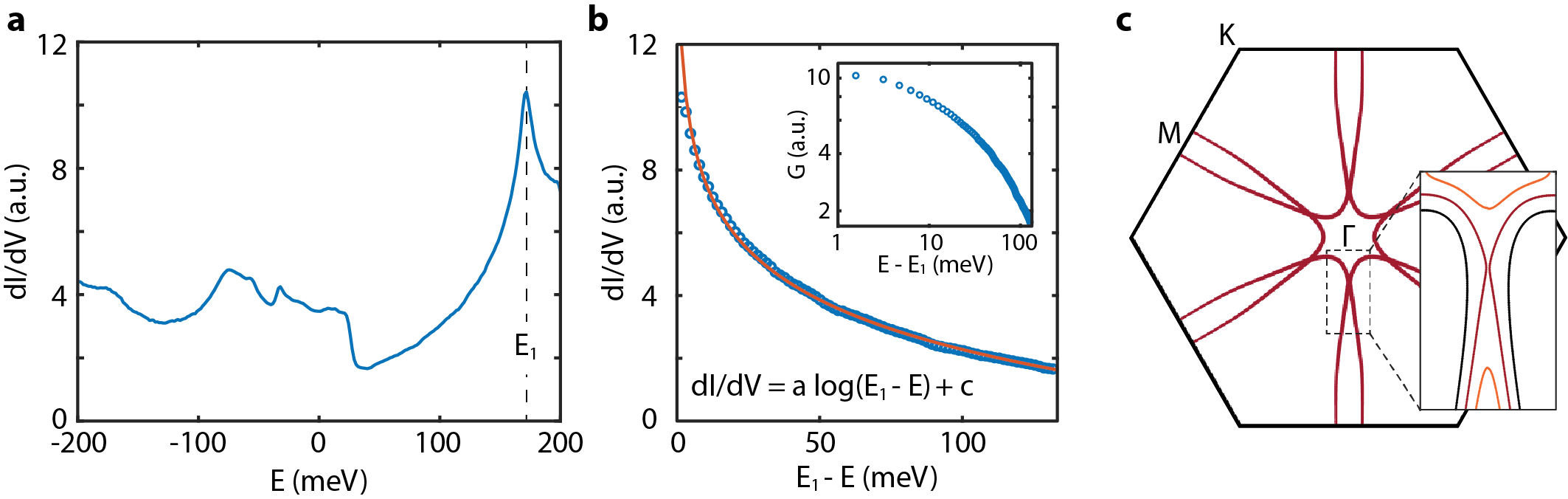} \caption{\textbf{Logarithmic
divergence in DOS on Bi(111).} \textbf{a,} $dI/dV$ profile measured on a
pristine Bi(111) surface showing the characteristic peak at $E_1$. \textbf{b,}
Zoomed in $dI/dV$ profile to the left of $E_1$ in \textbf{a} and a corresponding
logarithmic fit (orange). Inset shows the corresponding log-log plot.
\textbf{c,} Calculated band structure of Bi(111) surface in the vicinity of the
Lifshitz transition at about 180 meV ($E_1$). Inset: The zoomed in surface band
structure in orange, red and black correspond to 100 meV, 180 meV and 260 meV,
respectively.} \label{fig:Bi111} \end{figure}

We now focus on the energy region
just above the Dirac node in which the upper part of the Dirac band flattens out
as a function of energy. The $dI/dV$ spectrum in that energy window exhibits a
remarkable increase characterized by a sharp peak at $E_0=246$ meV
(Fig.\ref{fig:Bi110}d). This VHS is characterized by a power law divergence with
an unusually high exponent of $b=-0.70(2)$. Much higher than an exponent of
$0.25$ measured near high-order VHS in magic angle graphene \cite{Yuan2019a}.
Comparison with DFT calculations shows that the peak at $E_0=246$ meV is very
well captured by the calculated DOS stemming from the vicinity of the Dirac cone
at $\Gamma$ (Fig.\ref{fig:VHS}a and Fig.\ref{Sfig:DFT_DOS} for more details). A
power-law fit ($dI/dV \sim (E-E_0)^b$) to the $dI/dV$ profile shown in
Fig.\ref{fig:VHS}a, for energies above the peak, is presented in
Fig.\ref{fig:VHS}b along with a log-log plot of the exact fitting region
(inset). The power-law exponent was extracted by fitting over the linear region
shown in the log-log plot and its corresponding domain in the raw $dI/dV$
profile (dark blue circles in Fig.\ref{fig:VHS}b). This remarkable divergence
has been observed across all our measurements done at different regions on the
Bi(110) surface (see Fig.\ref{Sfig:VHS}). A similar analysis of the power-law
divergence of the DOS for energies below the peak is shown in
Fig.\ref{fig:VHS}c. The corresponding power-law fit yields a much lower exponent
of $b=-0.2$, possibly due to the DOS contributed by other bands, at lower energies. The high order
VHS observed on the (110) surface can be contrasted with a conventional VHS
observed on the (111) surface, at about 180 meV, as demonstrated by the $dI/dV$
spectrum of the Bi(111) surface shown in Fig.\ref{fig:Bi111}a and b. Here the
VHS originates from six ordinary saddle points that form around $\Gamma$ at 180
meV (Fig.\ref{fig:Bi111}c) when the six hole pockets along the $\Gamma - M$
direction merge with the hexagonal electron pocket around $\Gamma$ (see
Fig.\ref{Sfig:Bi111}). Indeed, the divergence exhibits logarithmic behaviour as
expected for conventional VHS (Fig.\ref{fig:Bi111}b).

Close inspection of the Bi(110) band structure around $\Gamma$, in the
corresponding energy window, reveals two Lifshitz transitions at slightly
different energies, $\sim 198$ meV and $\sim 229$ meV, which give rise to two
pairs of $C_2$ symmetric saddle points located along the $\Gamma - X_1$ and the $\Gamma - X_2$
directions, respectively (Fig.\ref{Sfig:CCE_Bi110}b and d. While the points
along $\Gamma - X_1$ are clearly ordinary saddle points, the nearly tangential
band touching along the $\Gamma - X_2$ direction (Fig.\ref{fig:VHS}d and e) have
similar shape to the $C_2$ symmetric VHS observed on twisted bilayer graphene
\cite{Yuan2019a} with a slight deviation just around the touching point. Indeed,
probing the DOS on a smaller region just around these nearly tangential touching
points does not account for the large exponent we observe. However, on a larger
scale (Fig.\ref{fig:VHS}a) the power-law exponent extracted on the Bi(110)
surface is significantly higher than the one observed on twisted bilayer
graphene \cite{Yuan2019a}. We therefore believe that the enhanced power-law
divergence does not originate from the immediate local surrounding of the saddle
points but rather from the unique topology of the Bi(110) band structure, at
this energy range, on a larger scale. As shown in Fig.\ref{fig:VHS}e and
Fig.\ref{Sfig:CCE_Bi110}, the surface bands along the $\Gamma - X_1$ direction
that constitute the nearly tangential saddle points remain flat almost across
the whole Brillouin zone (Fig.\ref{fig:VHS}f). At higher energies these flat
bands merge with the hole pockets along $\Gamma - X_2$, such that around the
peak energy $\sim 246$ meV, this surface band becomes flat along both the
$\Gamma - X_1$ and the $\Gamma - X_2$ directions (Fig.\ref{fig:VHS}e and
\ref{Sfig:CCE_Bi110}), though not to the same extent. This extended flattening
seems to give rise to the sharp increase in the DOS we observe in our
measurements. We therefore identify the small peak at $\sim 198$ meV with the regular saddle points along $\Gamma - X_1$, and the more significant peak with the extended flattening of the bands constituent the nearly tangential saddle points. While point singularities have been theoretically explored and
classified \cite{Chandrasekaran2020,Yuan2019a}, singularities of lines or
circles of critical points have not been classified and explored theoretically
\cite{Chandrasekaran2020}, further studies are needed to generalize the
classification to include singularities of infinite dimensions. The bulk chemical potential in Bi(110) lies around 200 mev above the touching point of the Dirac point and the higher order saddle point. Nevertheless, our spectroscopic study may very well motivate thin film growth and investigation \cite{yangg2020,laisichen2022} of this exotic orientation under efficient gate tuning of the electronic density towards the extreme van Hove singularities and adjacent Dirac node.

\section*{Summary}
In summary, we observe a power-law divergent DOS stemming from high-order VHS on
the (110) surface of Bismuth. This high order VHS does not originate from the
immediate local vicinity of the saddle points hosted on the surface, but rather
from an extended flattening of the surface band structure on a larger scale. We
further show that this high-order singularity occurs in close proximity to the
surface Dirac cone of Bi(110) surface. The coexistence of Dirac fermions and
nearly flat bands have been studied theoretically on 2D systems with
square-octagon lattice \cite{Oriekhov2021}. It was shown that the competition of
paramagnetic contribution of the high-order VHS and diamagnetic contribution of
the Dirac cones may lead to a dia- to paramagnetic phase transition in orbital
suseptibility as a function of doping at the touching point at which the
high-order VHS and Dirac point are present. Our observation renders Bi(110) an
interesting system to study physical quantities related to high-order VHS in
general and to this unique coexistence in particular. Moreover, tuning of the chemical potential towards the high-order VHS in Bi(110) thin films may induce correlated states of matter within strongly interacting topological bands.

\clearpage
\section*{Figures}

\begin{figure}[ht] \centering \includegraphics[width=\linewidth]{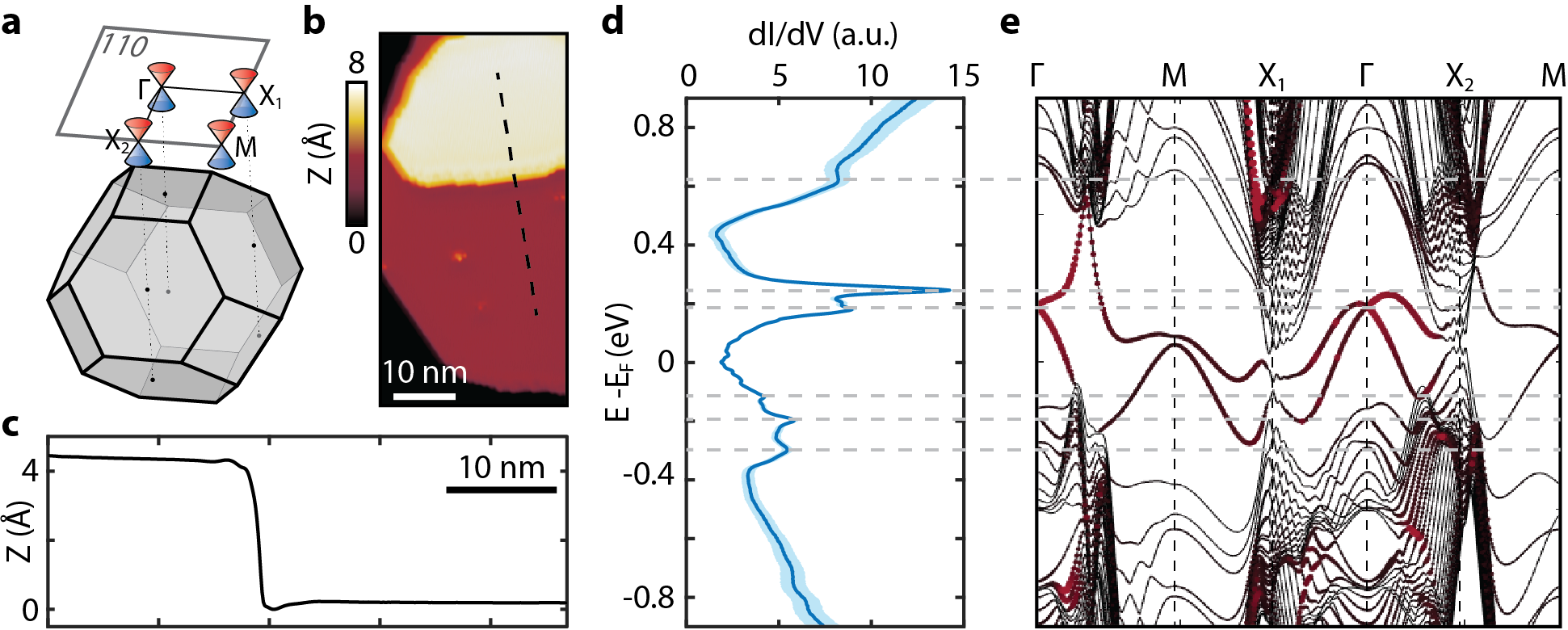}
\caption{\textbf{Electronic structure of Bi(110).} \textbf{a,} Schematic
illustration of the bulk Brillouin Zone and the projected 2D Brillouin Zone on
the 110 surface with Dirac cones at the TRIM points. \textbf{b,} Topography of
the Bi(110) surface. \textbf{c,} Height profile of an atomic step edge along the
dashed line marked in \textbf{b}. \textbf{d,} An average spectrum ($dI/dV$)
measured on the pristine Bi(110) surface (solid blue). The shaded region marks
the variation within 95\% confidence interval in the $dI/dV$ profile.
\textbf{e,} Calculated band structure of the Bi(110) surface along the high
symmetry lines. Additionally, small gaps of surface bands at $\Gamma$ and $M$
are artificially caused by the finite size effect of the slab model and should
vanish in reality. } \label{fig:Bi110} \end{figure}

\clearpage 

\clearpage 

\clearpage 

\clearpage \section*{Methods} \subsection*{STM measurements} The measurements
were performed in a commercial Unisoku STM at 4.2 K. The Pt-Ir tips were
characterized in a freshly prepared Cu(111) single crystal. This process ensured
a robust tip with reproducible results across different cleaves and samples from
different batches. All the $dI/dV$ measurements were taken using standard
lock-in techniques.

\subsection*{DFT calculations} To study the electronic properties of Bi(110)
surface, $ab~initio$ calculations based on density functional theory (DFT) were
performed using Vienna Ab initio Simulation Package
(VASP)\cite{Kresse1996,Tran2009}. The projector-augmented wave pseudopotential
and a plane-wave energy cutoff of 120 eV were adopted. A 58 \AA~ thick slab was
constructed to model the cleaved Bi(110) surface. The in-plane lattice
parameters of 4.25 ×4.72 \AA~ and a vacuum region of $\geq$ 15 \AA~ along
z-direction were set for the slab model. A $12\times12\times1$ k-mesh was
adopted for sampling the two-dimensional Brillouin Zone. The four upmost atomic
layers were relaxed until the force on each atom is less than 0.01 eV/\AA. The
surface-atom-projected band structures in Fig.\ref{fig:Bi110}e present the
surface electronic properties. The size of red circules represent the projected
weight of the top four Bi layers. Because of the finite size effect, the top and
bottom surface states have weak hybridization and generate tiny gaps at $\Gamma$
and $M$ points. Such gaps have marginal effects in our analysis of VHS and
should vanish in reality.

\section*{Data availability} The data that support the plots within this paper
and other findings of this study are available from the corresponding authors
upon reasonable request.

%\clearpage

\printbibliography[title=References]

\section*{Acknowledgement} The authors thank A. Stern and R. Quieroz for
insightful discussions. N.A., H.B., and B.Y acknowledge the German–Israeli
Foundation for Scientific Research and Development (GIF grant no.
I-1364-303.7/2016). B.Y. acknowledges financial support by the Willner Family
Leadership Institute for the Weizmann Institute of Science, the Benoziyo
Endowment Fund for the Advancement of Science, the Ruth and Herman Albert
Scholars Program for New Scientists, and the Israel Science Foundation (ISF
1251/19). T.P. acknowledges financial support from NSERC and FRQNT.

\section*{Author contribution} A.K.N. and J.R. acquired the data. A.K.N analyzed
the data. A.K.N, N.A and H.B. conceived the experiments. H.T., H.F. and B.Y.
calculated the ab initio model. H.L., and T.P. calculated the theoretical model.
C.S. and C.F. grew the material. A.K.N., H.B., and N.A. wrote the manuscript
with substantial contributions from all authors.

\section*{Competing interests}
The authors declare no competing interests.

\include{SM/SM}

\end{document}

%% file: SM/SM.tex
%% Figure caption
\renewcommand{\figurename}{\textbf{Fig.}}
\renewcommand{\thefigure}{S\arabic{figure}}
\setcounter{figure}{0}  

%% Section
\renewcommand{\thesection}{S\arabic{section}}

\title{Supplementary Information for \\ Visualizing a Dirac cone in proximity to high-order van Hove singularities}

\maketitle

\begin{refsection}

\section{Additional QPI measurements}
\begin{figure}[ht]
\centering
\includegraphics[scale=1]{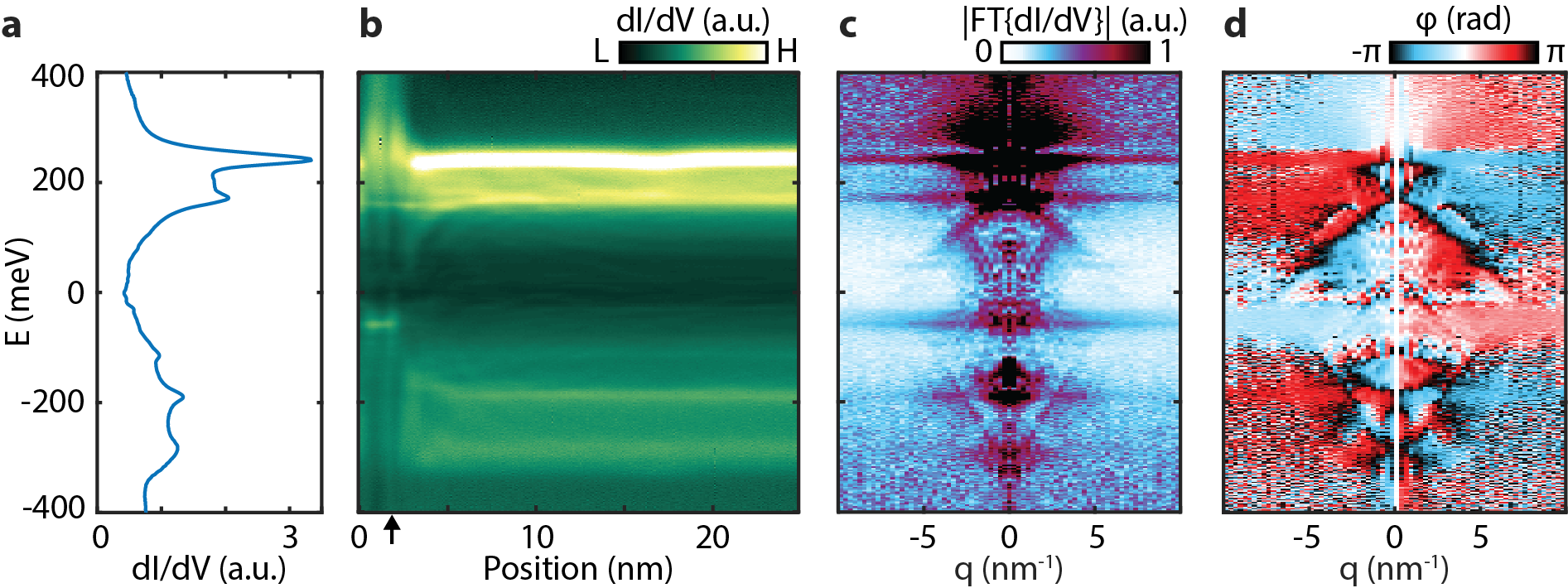}
\caption{\textbf{$dI/dV$ map on Bi(110) sample \#1.} \textbf{a,} $dI/dV$ profile measured far from the step edge. \textbf{b,} $dI/dV$ mapping of the QPI patterns due to scattering off the step edge (marked by black arrow). \textbf{c,d,} The magnitude (same as Fig.3c) and the phase ($\phi$) of the Fourier transform of \textbf{b} showing the various dispersing modes, respectively.}
\label{Sfig:QPI_2}%https://www.overleaf.com/project/5fe452b7634e2803073a5daf
\end{figure}

\begin{figure}[ht]
\centering
\includegraphics[scale=1]{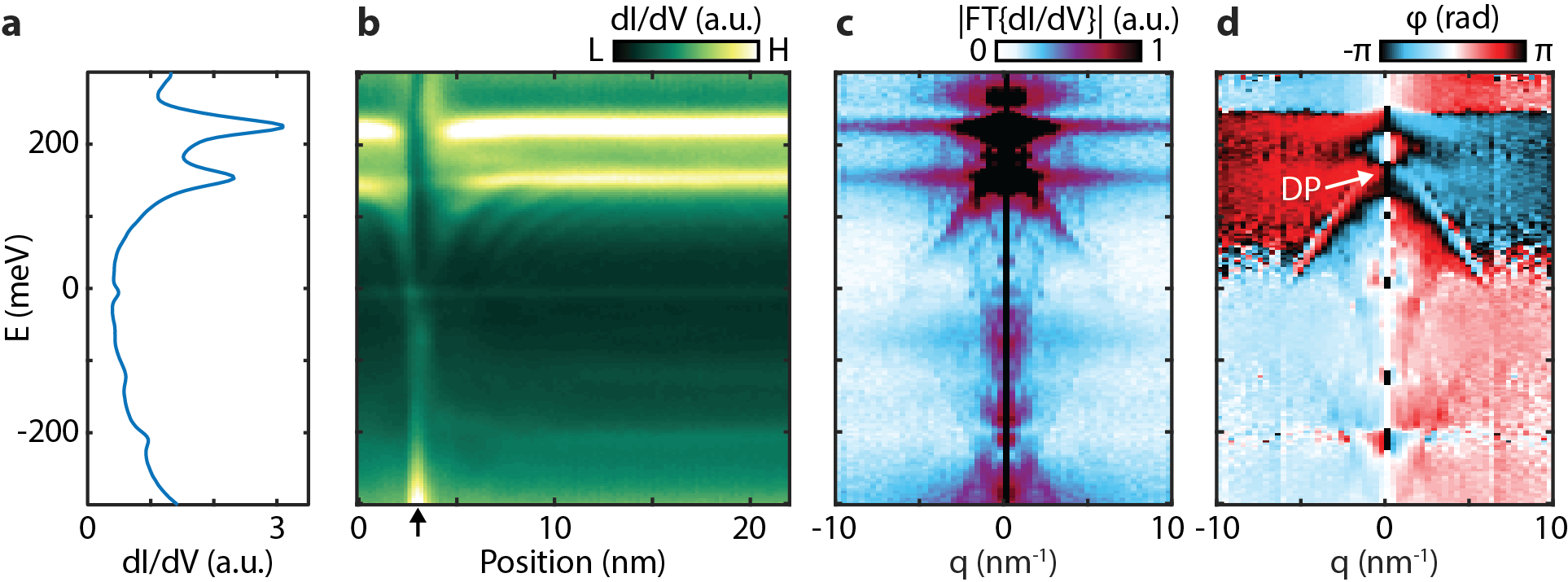}
\caption{\textbf{$dI/dV$ map on Bi(110) sample \#2.} Similar measurement as Fig.\ref{Sfig:QPI_2} for sample \#2.}
\label{Sfig:QPI_S2}
\end{figure}

Similar QPI patterns were observed in the dI/dV measurements on two different samples as shown in Fig.\ref{Sfig:QPI_2} and \ref{Sfig:QPI_S2}, respectively. The dI/dV profile measured far from any defects and impurities on the (110) surface is shown in Fig.\ref{Sfig:QPI_2}a. The measured profile shows the characteristic peak at $E\sim 240$ meV. The dI/dV map measured perpendicular to a step edge (marked by black arrow) is shown in Fig.\ref{Sfig:QPI_2}b. It has dispersing features emanating from the step edge and non-dispersing features associated with the surface band extrema and van Hove singularities. Fourier transform of the dI/dV map shows the evolution of several dispersing features. The magnitude and the phase of the Fourier transform is shown in Fig.\ref{Sfig:QPI_2}c and d, respectively. Most of the features are well identified in the calculated QPI. The phase of the QPI pattern may reveal information related to the Berry curvature of the surface bands \cite{Dutreix2019,Zhang2020}. However, here we use the phase as a means to clearly visualize the scattering processes. Similar QPI features can be seen in another measurement on a different sample, as shown in \ref{Sfig:QPI_S2}.

\section{Identification of scattering processes}
\begin{figure}[ht]
\centering
\includegraphics[width=\textwidth]{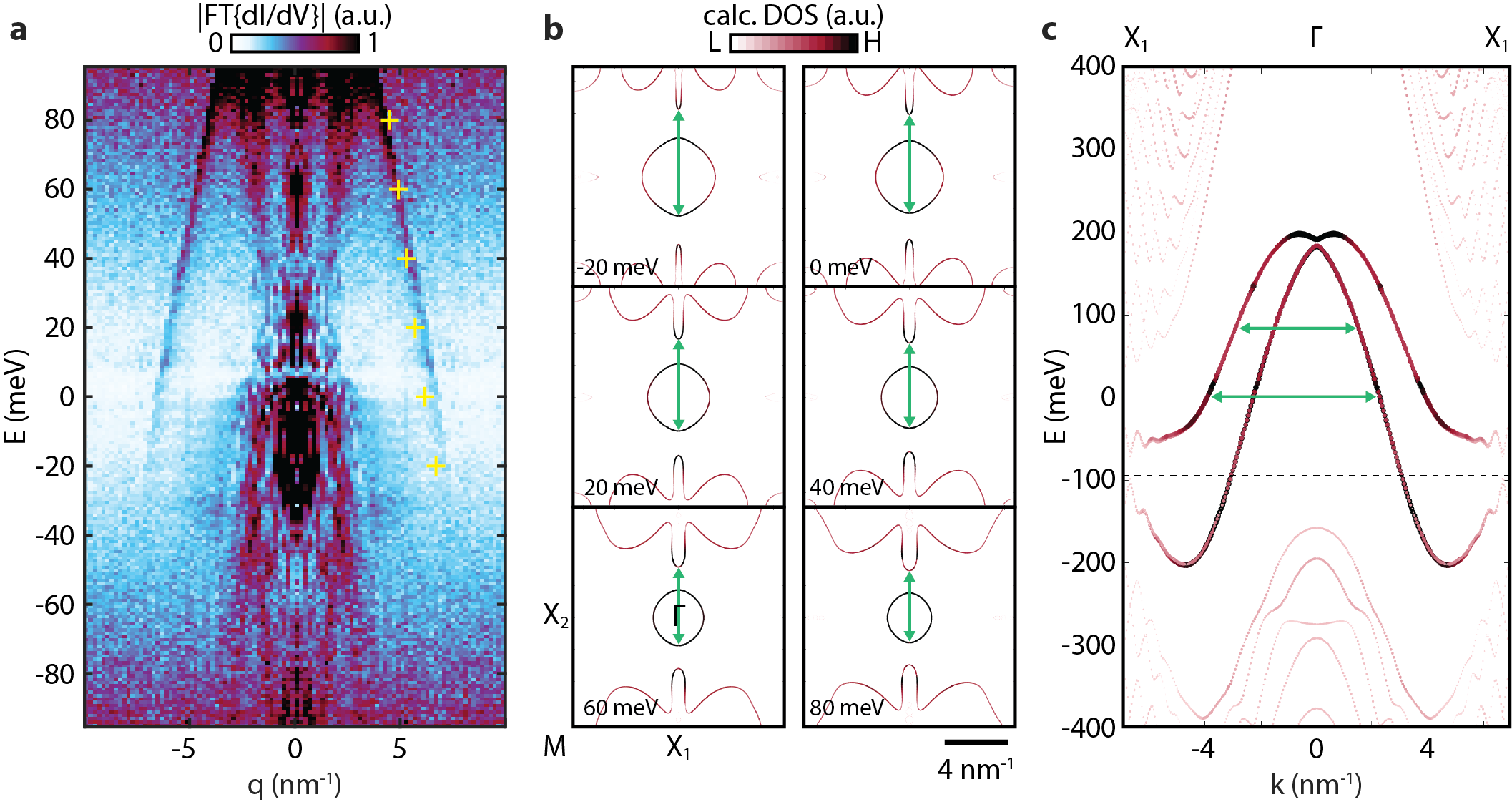}
\caption{\textbf{Scattering across Dirac bands.} \textbf{a,} Fourier transform of the measured $dI/dV$ profile on Bi(110) surface (same as Fig.\ref{fig:Bi110}b). \textbf{b,} Calculated CCEs for the Bi(110) surface. The scattering process at various energies is marked by the green arrows. The corresponding \textbf{q} values are plotted in \textbf{a} as yellow '+'. \textbf{c,} Calculated band structure of Bi(110) surface along the high-symmetry direction $X_1-\Gamma-X_1$.}
\label{Sfig:SP1}
\end{figure}

\begin{figure}[ht]
\centering
\includegraphics[scale=1]{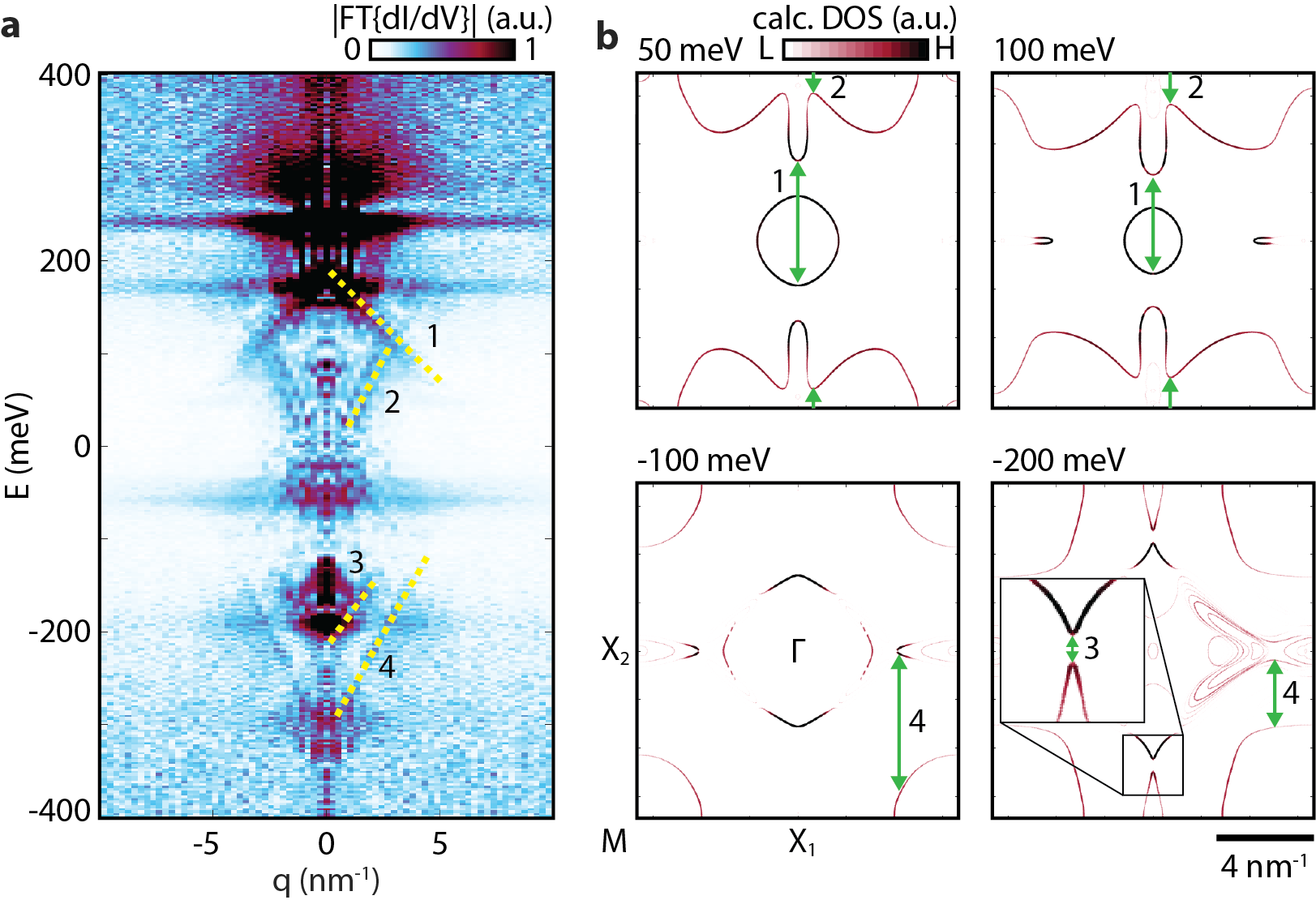}
\caption{\textbf{Various scattering wave vectors.} \textbf{a,} Fourier transform of dI/dV measurement (same as Fig.3c). \textbf{b,} Scattering wave vectors are labeled and marked (green arrows) on some of the representative CCEs of the Bi(110) surface band structure.}
\label{Sfig:SP2}
\end{figure}

\begin{figure}[ht]
\centering
\includegraphics[scale=1]{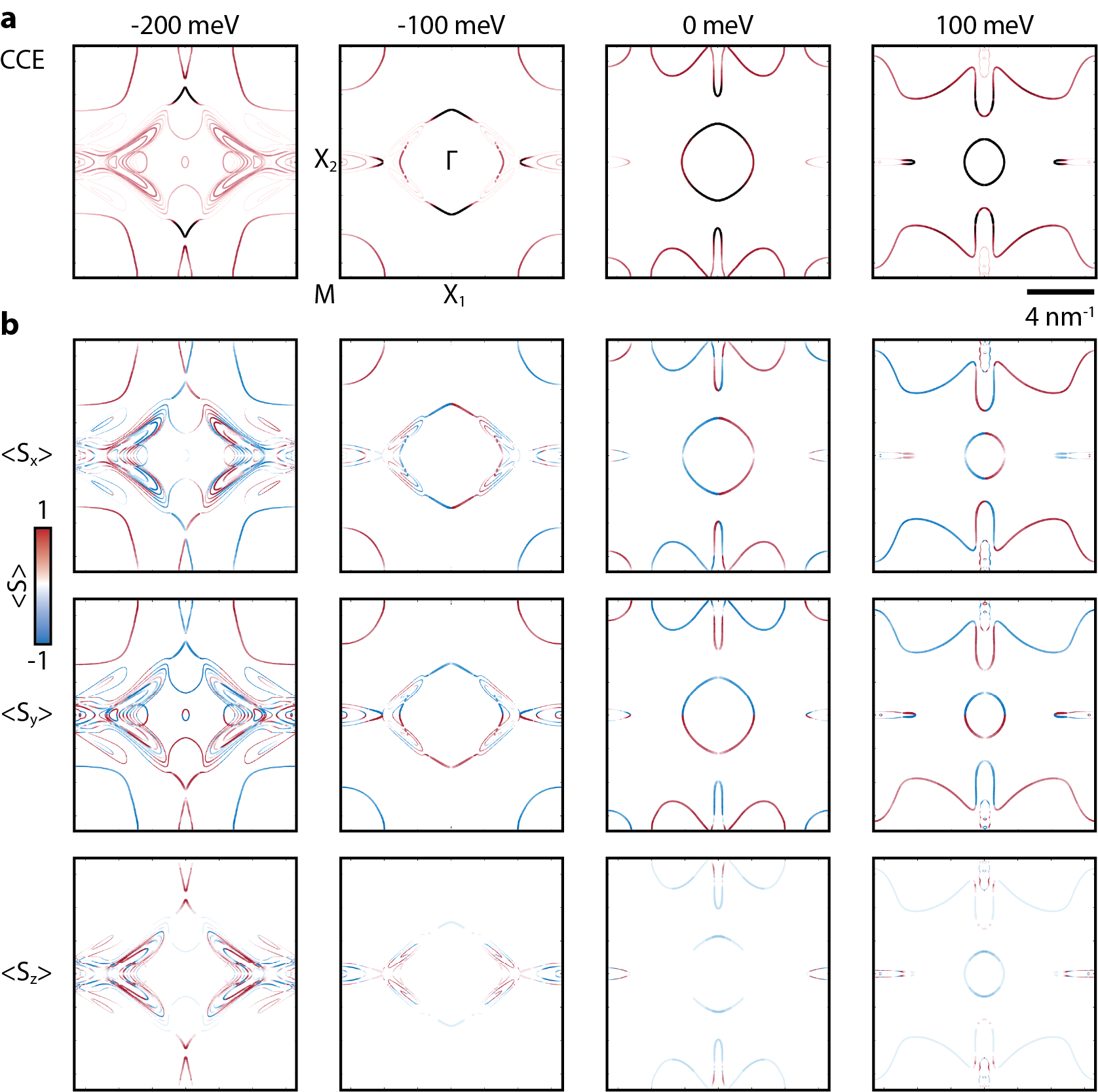}
\caption{\textbf{Calculated spin texture of the Bi(110) surface.} \textbf{a,} CCEs of the Bi(110) band structure at representative energies. \textbf{b,} Normalized spin components of the corresponding CCEs in \textbf{a}.}
\label{Sfig:spin}
\end{figure}

The detailed scattering processes giving rise to the observed QPI patterns is presented in Fig.\ref{Sfig:SP1} and Fig.\ref{Sfig:SP2}. Fig.\ref{Sfig:SP1}a shows a high energy resolution QPI map of Bi(110) surface. We clearly identify the main scattering process observed in the QPI map as the intra-Dirac scattering along $X_1-\Gamma-X_1$, as shown in Fig.\ref{Sfig:SP1}b and c. Such scattering processes are not forbidden since it occurs across the same Dirac band with a finite overlap in their spin texture (Fig.\ref{Sfig:spin}). Similarly, more scattering process are identified in the large energy window QPI map, as shown in Fig.\ref{Sfig:SP2}.

\section{Calculated QPI pattern}
In order to calculate the QPI pattern we use the Born approximation which states that
\begin{eqnarray}
\delta N(q,\omega) = -{\frac{1}{\pi}}V(q){\rm Im}\left( \int {\frac {d^d k}{(2\pi)^d}} G(k,\omega) G(k-q,\omega) \right).
\end{eqnarray}

In other words, the spatial modulations in the density of states at energy $\omega$, $\delta N(r,\omega)$ is a consequence of scattering events due to the potential $V(r)$.  When Fourier transformed, the local density of states is arranged according to the momentum transfer $q$ and hence it is a convolution of the Green's function with itself.  This convolution relates on-shell states with energy $\omega$ (where the Green's functions are peaked) such that the initial and final momenta are separated by the vector $q$.

In order to construct the Green's functions we use the wavefunctions which are found by DFT to have the largest weight on the surface and write the retarded Green's function as:
\begin{eqnarray}
G(k,\omega) = \sum_n{\frac{|\Psi_{k,n}\rangle \langle \Psi_{k,n}|}{\omega -\epsilon_{k,n}+i\eta}}
\end{eqnarray}
where $n$ goes over all relevant eigen modes, $\epsilon_{k,n}$ is its corresponding energy and $\eta$ is a small phenomenological broadening of the states.  $\Psi_{k,n}$ is a spinor whose direction is found by DFT.
\begin{figure}[ht]
\centering
\includegraphics[scale=1]{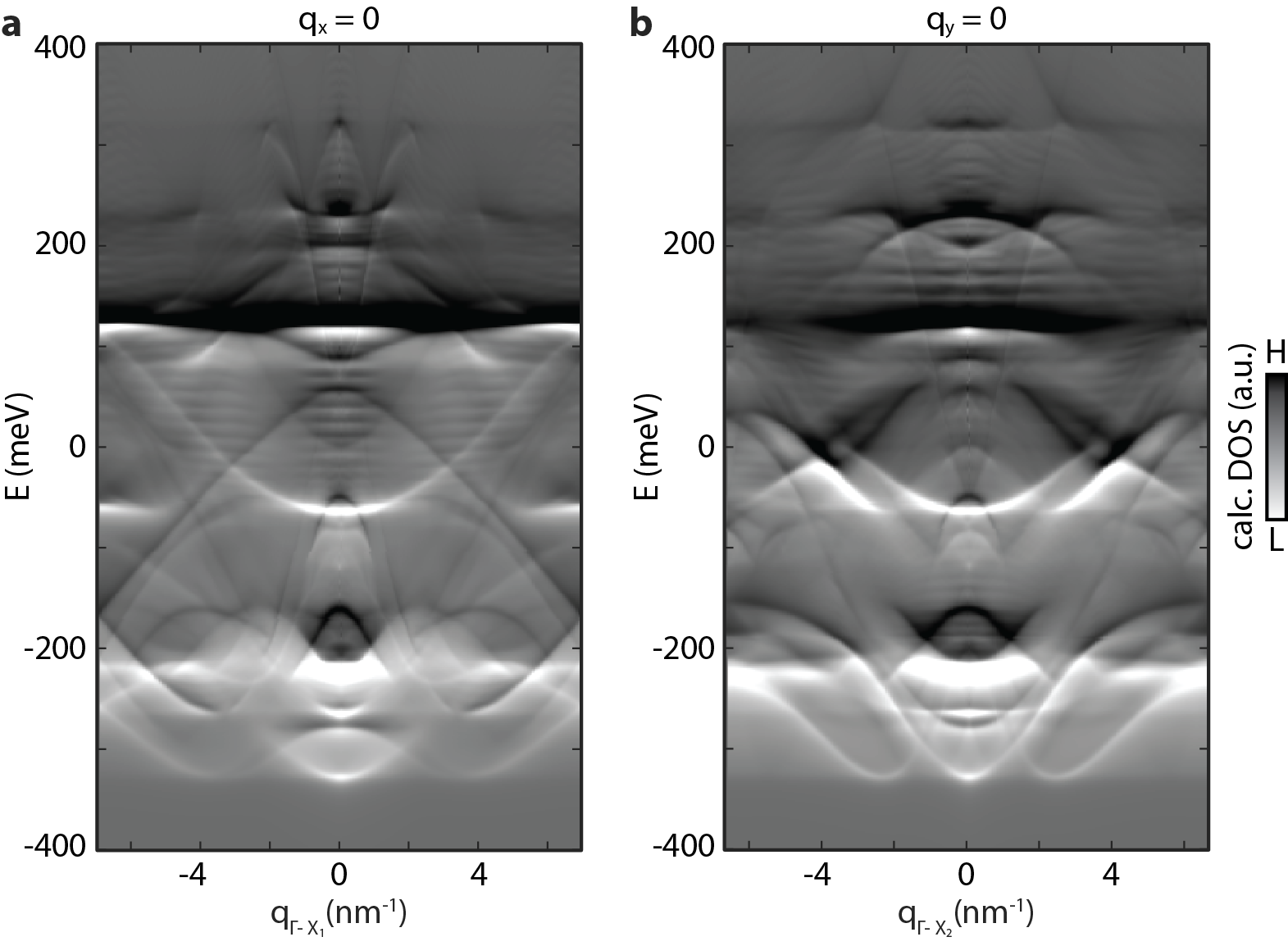}
\caption{\textbf{Calculated QPI using Green's function approach.} \textbf{a,b,} Calculated QPI cut along $\Gamma-X_1$ and $\Gamma-X_2$ direction, respectively.}
\label{Sfig:GF}
\end{figure}

\section{DOS of Bi(111)}
\begin{figure}[ht]
\centering
\includegraphics[scale=1]{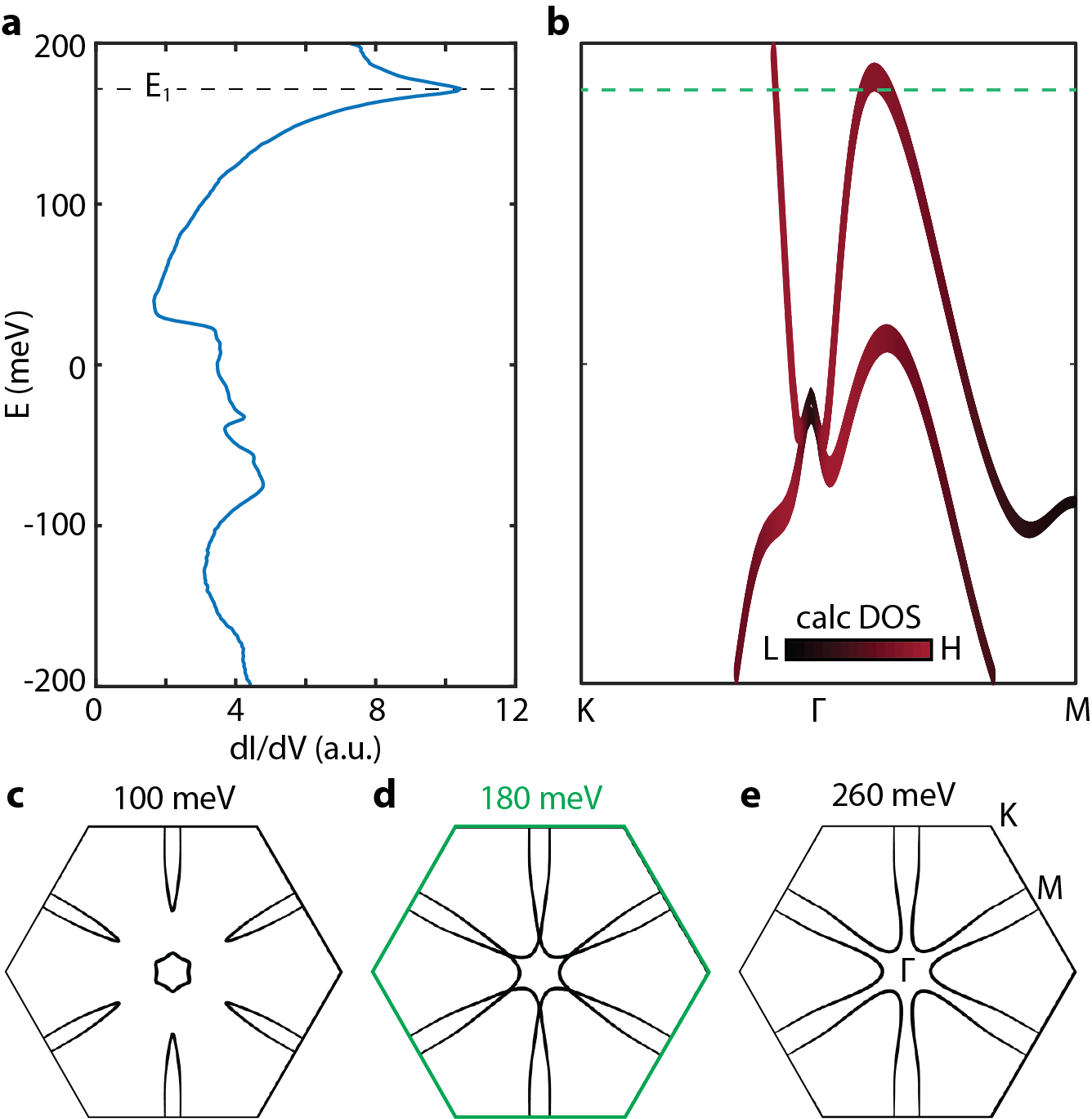}
\caption{\textbf{Calculated band structure of Bi(111) surface.} \textbf{a,} A typical $dI/dV$ profile measured on a clean (111) surface (same as Fig.2f). \textbf{b,} Ab initio calculated band structure of the Bi(111) surface along a high symmetry line. \textbf{c-e,} Contours of constant energy showing the evolution of the bands in the vicinity of the Lifshitz transition (green) responsible for the $E_1$ peak.}
\label{Sfig:Bi111}
\end{figure}

The DOS measured on a prisitne Bi(111) surface is shown in Fig.\ref{Sfig:Bi111}a (more details in ref.\cite{Nayak2019}). It shows a characteristic peak at $E_1 \sim 180$ meV. This peak corresponds to the band extrema along the $\Gamma - M$ direction as shown in Fig.\ref{Sfig:Bi111}b. The DOS diverges logarithmically close to the peak $E_1$ due to ordinary van-Hove singularities (Fig.\ref{Sfig:Bi111}c-e). The topology of the Fermi surface undergoes a transition at the peak energy. This demonstrates the ability to distinguish ordinary and high-order van Hove singularities in dI/dV profiles. 

\section{Calculated DOS on Bi(110)}
\begin{figure}[ht]
\centering
\includegraphics[width=\textwidth]{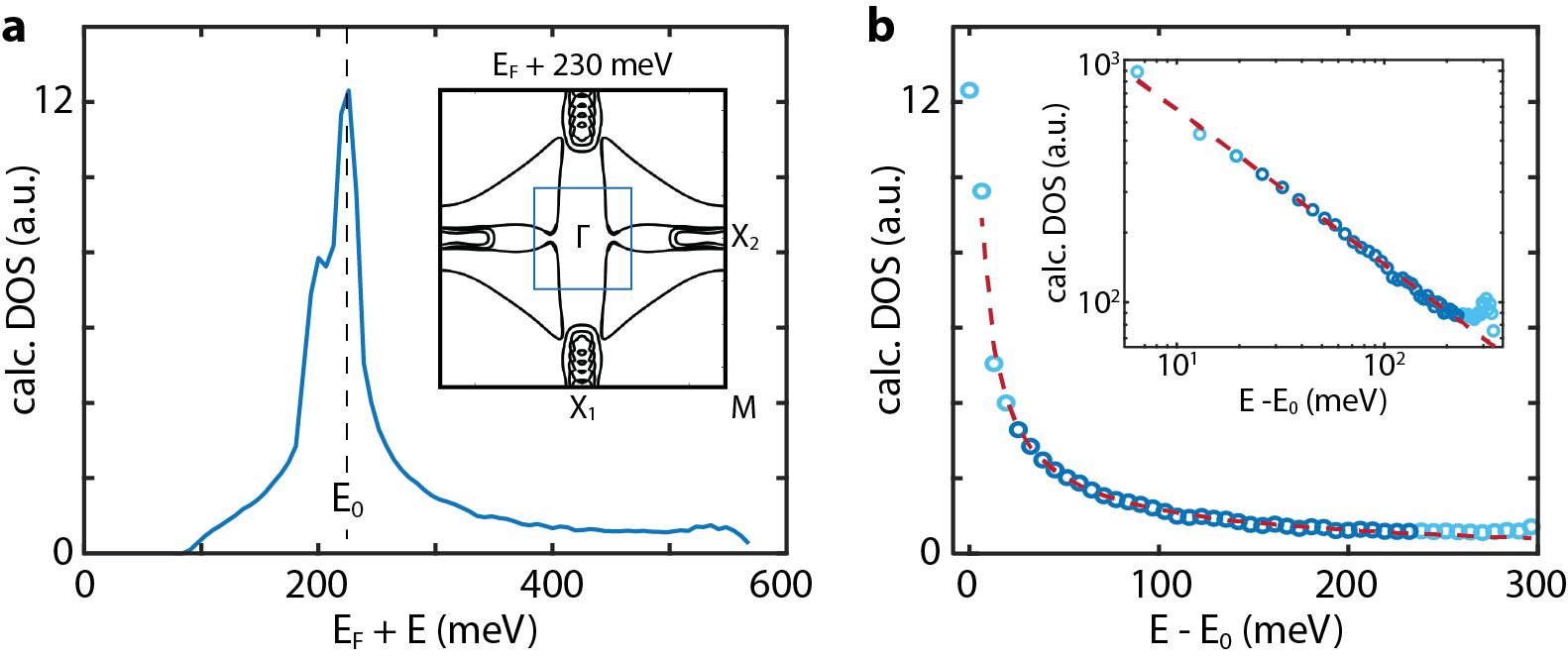}
\caption{\textbf{DFT calculated DOS on Bi(110).} \textbf{a,} Momentum integrated DOS extracted from DFT surface band structure in the vicinity of the zone center (marked by blue box in inset). Inset: Contour of constant energy at 230 meV above the Fermi energy. \textbf{b,} Power-law fit to the calculated DOS, $a(E-E_0)^b+c$. The fit yields an exponent of $b=0.67(3)$. Inset shows linear fit to log-log plot, $m(E-E_0)+c$. The fit yields a slope of $m=0.67(2)$.}
\label{Sfig:DFT_DOS}
\end{figure}

We extracted the total DOS from Density Functional Theory (DFT) calculated surface band structure. The calculated (calc.) DOS was integrated only over a region close to the $\Gamma$ point as shown in the inset of Fig.\ref{Sfig:DFT_DOS}a. This region (marked by a blue box) includes the high-order van-Hove singularity close to the $\Gamma$ point. The calculated DOS to the right of the peak $E_0$ fits well to a power-law model as shown in Fig.\ref{Sfig:DFT_DOS}b. The same data was plotted in log-log to unambiguously determine its power-law nature (inset Fig.Fig.\ref{Sfig:DFT_DOS}b).

\section{Power-law divergent DOS}
\begin{figure}[ht]
\centering
\includegraphics[width=\textwidth]{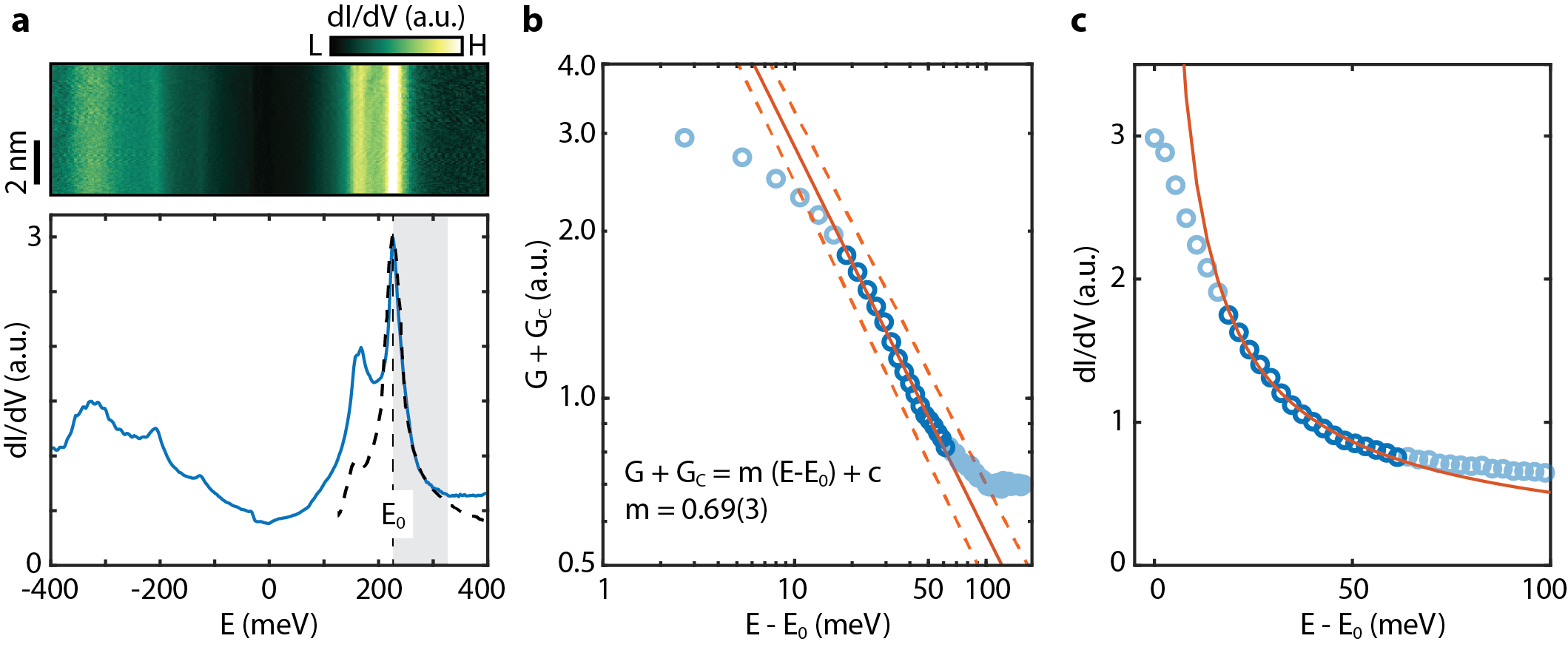}
\caption{\textbf{Diverging DOS on the Bi(110) surface.} \textbf{a,} dI/dV map on Bi(110) away from defects (upper panel) and spatially averaged dI/dV profile overlaid with calculated DOS (lower panel). \textbf{b-c,} Log-log and linear plot of the dI/dV profile close to the diverging peak ($E_0$) marked by the shaded region in \textbf{a}. Linear and corresponding power-law fit to the density of states is shown in \textbf{b} and \textbf{c}, respectively.}
\label{Sfig:VHS}
\end{figure}

The power-law divergence in the density of states was carefully measured across measurements on different regions of the sample as shown in Fig.2 and \ref{Sfig:VHS}. The spatially averaged dI/dV profile measured on the Bi(110) surface is shown in  Fig.\ref{Sfig:VHS}a (lower panel). The dI/dV profile was averaged over a clean region on the (110) terrace (upper panel Fig.\ref{Sfig:VHS}a). The dI/dV profile shows a characteristic peak, marked by $E_0$, corresponding to van Hove singularity in the surface band structure. The \textit{ab initio} calculated total DOS (black dashed line), overlaid on the dI/dV profile, shows good agreement (see Fig.\ref{Sfig:DFT_DOS} for details). We examined the power-law behavior in the density of states in the shaded region of Fig.\ref{Sfig:VHS}a.

First, we plot the dI/dV profile in log-log to clearly identify the linear regime (marked by dark blue circles in Fig.\ref{Sfig:VHS}b). We fit this subset of the dI/dV profile (that is linear in log-log plot) to a power-law formula $a(E-E_0)^b+d$. This fit yields the following parameters: $a=14(6)$, $b=-0.7(2)$, and $d=-0.06(30)$. To accurately extract the slope from the log-log plot, we add the constant $d$ extracted from the power-law fit to the DOS in the log-log plot. Therefore, $G + G_C$ is equivalent to $\log{\left(dI/dV - d\right)}$ in Fig.\ref{Sfig:VHS}b. Fitting a linear model, $m(E-E_0)+c$, to the log-log plot over the relevant energy window (marked by dark blue circles) yields the slope $m=0.7(2)$, consistent with the power-law fit. The dI/dV profile close to the peak ($|E-E_0| \leq \sim 10$ meV) deviates from the power-law behavior, possibly due to instrumental broadening of the divergence.

\section{CCEs of Bi(110)}
\begin{figure}[ht]
\centering
\includegraphics[scale=1]{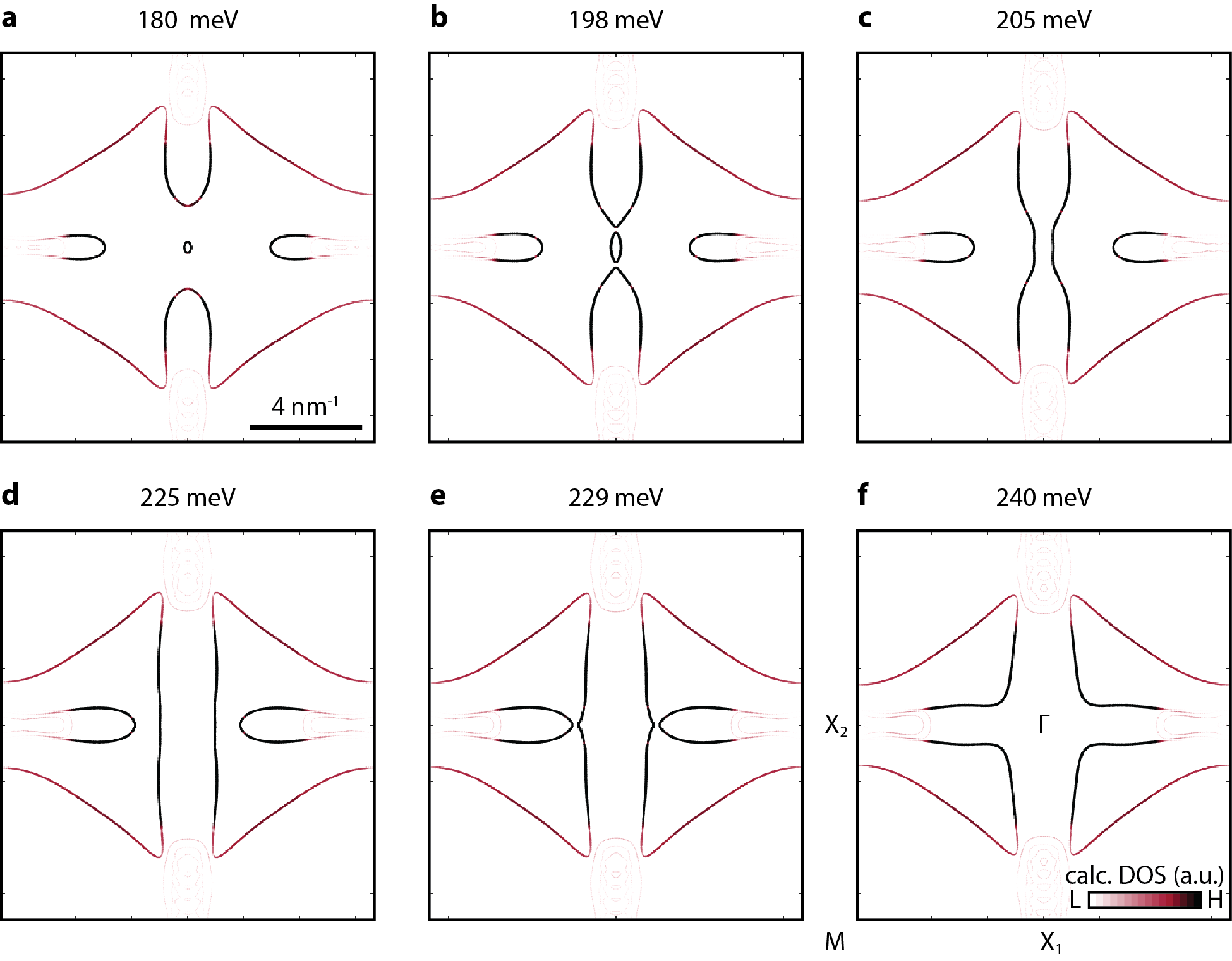}
Bismuth\caption{\textbf{Calculated band structure of Bi(110) surface.} Evolution of the Bi(110) surface band structure across Lifshitz transition in \textbf{b} and \textbf{e}. The nearly tangential band touching in \textbf{e} may give rise to high-order van Hove singularity.}
\label{Sfig:CCE_Bi110}
\end{figure}

The contours of constant energy (CCE) calculated for the Bi(110) surface is shown in Fig.\ref{Sfig:CCE_Bi110}. The CCEs show the transition in the Fermi pockets known as Lifshtiz transition. The first saddle point appears along $\Gamma-X_1$ at $E=198$ meV and the second saddle point appears along $\Gamma-X_2$ at $E=229$ meV. The Dirac cone at $\Gamma$ merges with the pockets along $\Gamma-X_1$ (Fig.\ref{Sfig:CCE_Bi110}b) creating long straight segments in the CCE as shown in Fig.\ref{Sfig:CCE_Bi110}d. Subsequently, the long straight segments merge with the pockets along $\Gamma-X_2$ almost tangentially as shown in Fig.\ref{Sfig:CCE_Bi110}e.

\clearpage
\printbibliography[title=Supplementary References]

\end{refsection}